\documentclass[draft,ras]{agutex}

\emergencystretch=\maxdimen
  \usepackage[dvips]{graphicx}
  \usepackage{ulem}
  \usepackage{amssymb}
  \setkeys{Gin}{draft=false}
\authorrunninghead{LI ET AL.}

\titlerunninghead{THE FAST PROJECT}

\authoraddr{Corresponding author: D. Li,
National Astronomical Observatories,
Chinese Academy of Sciences, Beijing, China.
(dili@nao.cas.ac)}

\begin{document}

%
%

\title{The Five-hundred-meter Aperture Spherical Radio Telescope (FAST) Project}
%
%

%
%



\authors{Di Li\altaffilmark{1,2} and Zhichen Pan\altaffilmark{1}}

\altaffiltext{1}{National Astronomical Observatories,
Chinese Academy of Sciences, Beijing, China. dili@nao.cas.ac}

\altaffiltext{2}{Key Laboratory of Radio Astronomy,
Chinese Academy of Sciences, China.}

%
%


\begin{abstract}
The Five-hundred-meter Aperture Spherical radio Telescope (FAST) is a Chinese mega-science project funded by the National Development and Reform Commission (NDRC) of the People's Republic of China.
The National Astronomical Observatories of China (NAOC) is in charge of its construction and subsequent operation.
Upon its expected completion in September 2016,
FAST will surpass the 300-meter Arecibo Telescope and the 100-meter Green Bank Telescope in terms of absolute sensitivity in the 70 MHz to 3 GHz bands.
In this paper, we report on the project, its current status, the key science goals, and plans for early science.
\end{abstract}

%
%

%

\begin{article}

%
%


\section{The FAST Project}

\begin{figure}
\begin{center}
 \includegraphics{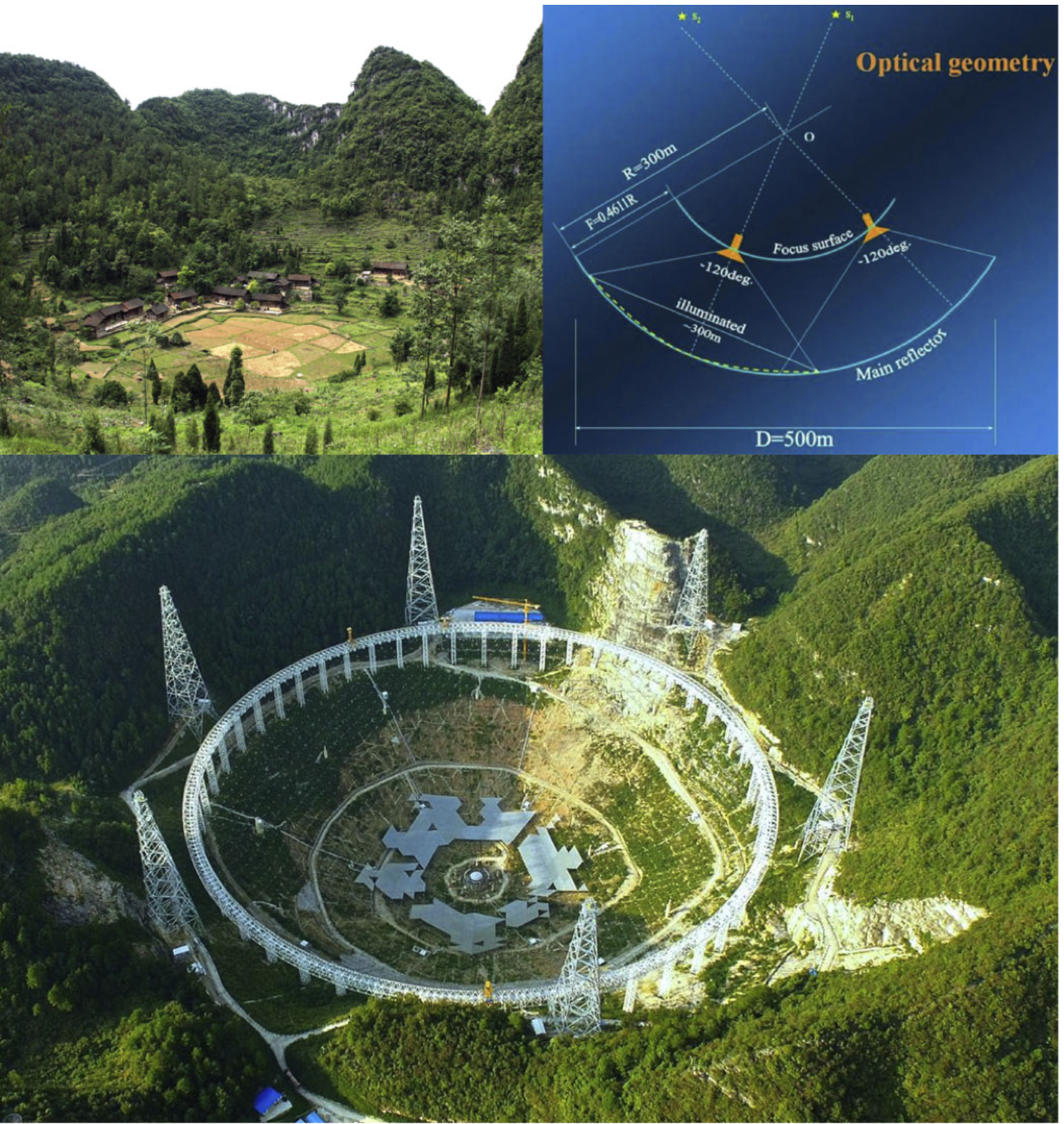}
 \caption{Upper Left: Partial view of the original site in November of 2009.
 Upper Right: FAST optical geometry.
 Lower: An aerial view of FAST under construction on September 26th, 2015 (credit: Shuxin Zhang of NAOC).}
   \label{fig1}
\end{center}
\end{figure}

As a Chinese Mega-Science program and contribution to the international effort to build the Square Kilometer Array (SKA),
the Five-hundred-meter Aperture Spherical radio Telescope (FAST: Nan et al. 2011)
started as part of an SKA concept and evolved into a standalone project.
An image of the original site, the schematics of the FAST optics, and a photograph of FAST under construction are shown in Fig. \ref{fig1}.
Upon its expected first light in 2016,
FAST will achieve the highest sensitivity within its frequency bands of any single-dish radio telescopes via three key advantages/innovations;
1) a site with deep karst terrain at Dawodang, in Guizhou province in southwestern China;
2) an active primary surface, 300 m in diameter, comprised of 4500 panels;
3) a light-weight feed cabin driven by six cables and servomechanisms plus a parallel robot realizing half-closed-loop precision control.
These features will provide almost twice the collecting area of the post-Gregorian upgrade Arecibo telescope (Altschuler 2002),
and almost ten times that of the Green Bank Telescope (GBT, Prestage et al. 2009) and the Effelsberg 100-m Telescope (Hachenberg, Grahl \& Wielebinski 1973).
It is also comparable to other SKA precursors,
such as ASKAP (Australian Square Kilometre Array Pathfinder, collecting area $\rm \sim$4000 m$^2$, covering 0.7 to 1.8 GHz, now working as an array with six antennae)
and MeerKAT (collecting area $\rm \sim$9000 m$^2$, covering 1.00 to 1.75 GHz), and the current SKA Phase I plan (SKA I Low: collecting area $\rm \sim$400000 m$^2$, covering 50 to 350 MHz; SLA I Mid: collecting area $\rm \sim$33000 m$^2$, covering 0.35 to 4.00 GHz.).
FAST will have a continuous frequency coverage between 70 MHz and 3 GHz and a maximum zenith angle of 40$^{\circ}$.
The raw sensitivity in L-band (1 to 2 GHz), which is the core band for FAST science, reaches 2000 m$^{2}$ K$^{-1}$.
When observing, the time to slew between sources is less than 10 min, given a maximum slew of 80$^{\circ}$.
A summary of the technical specifications of FAST can be found in Table 1.

\begin{table}
  \begin{center}
  \caption{Main technical specifications of FAST.}
  \label{tab1}
  \begin{tabular}{|l|}
  \hline
The longitude and the latitude: East 107$^{\circ}$21$¡ä$,  North 25$^{\circ}$48$¡ä$ \\
Spherical reflector: radius of curvature = 300 m, aperture diameter = 500 m \\
Illuminated parabolic primary aperture: Diameter = 300m  Area = $\rm \sim$7$\times$10$^4$ m$^{2}$ \\
Focal ratio: f/D = 0.4611  \\
Sky coverage: zenith angle 40$^{\circ}$ (Dec -14$^{\circ}$12$'$ to 65$^{\circ}$48$'$) \\
Frequency coverage: 70 MHz - 3 GHz  \\
Sensitivity (L-Band): A/T $\sim$2000 m$^{2}$ K$^{-1}$ ($\rm \sim$18 K Jy$^{-1}$), system temperature T$_s$$_y$$_s$ $\sim$20 K  \\
Beam FWHM (L-Band): 2.9$'$  \\
Field of view (L-band 19-beam feed-horn array): Diameter = 26$'$ \\
Slewing time: 1.5$-$10 minutes  \\
Pointing accuracy: 8$''$  \\
  \hline
  \end{tabular}
 \end{center}
\end{table}

The Chinese Astronomy community will strive to keep FAST open for international research proposals.
We will start system integration in June 2016.
Since March 2011, the FAST construction project has been proceeding on six fronts.

\textbf{1) Site Survey and Excavation.}
The natural shape of the karst terrain at the site is within a few percent of the desired spherical end-cap in terms of volume within 500 m diameter.
Nevertheless, we still removed about one million cubic meters of earth.
The site excavation, reformation, and protection against the fall of dangerous rocks are now complete.

\textbf{2) Active Reflector.}
The FAST primary surface comprises,
(i) 4300 triangular panels with varying sizes of about 11 m per side,
(ii) 150 square panels with a range of sizes filling the edges of the dish up to a girder ring (the white circular structure in the lower panel of Figure 1) of exactly 500 m diameter (Jiang, Wang, \& Zhao 2003).
In order to avoid spherical aberration, a parabolic surface is needed for single point focussing.
Through tie-down cables (see details in Nan et al. 2011 and Jiang et al. 2015),
the panels are driven in real time by 2225 hydraulic actuators to form a 300 m diameter parabolic surface out of a portion of the originally spherical surface whose radius of curvature is 300 m (see Figure 1).
Along the symmetry axis, the focal point is then about 4 m above the plane defined by the girder ring.

\begin{figure}
\begin{center}
 \includegraphics{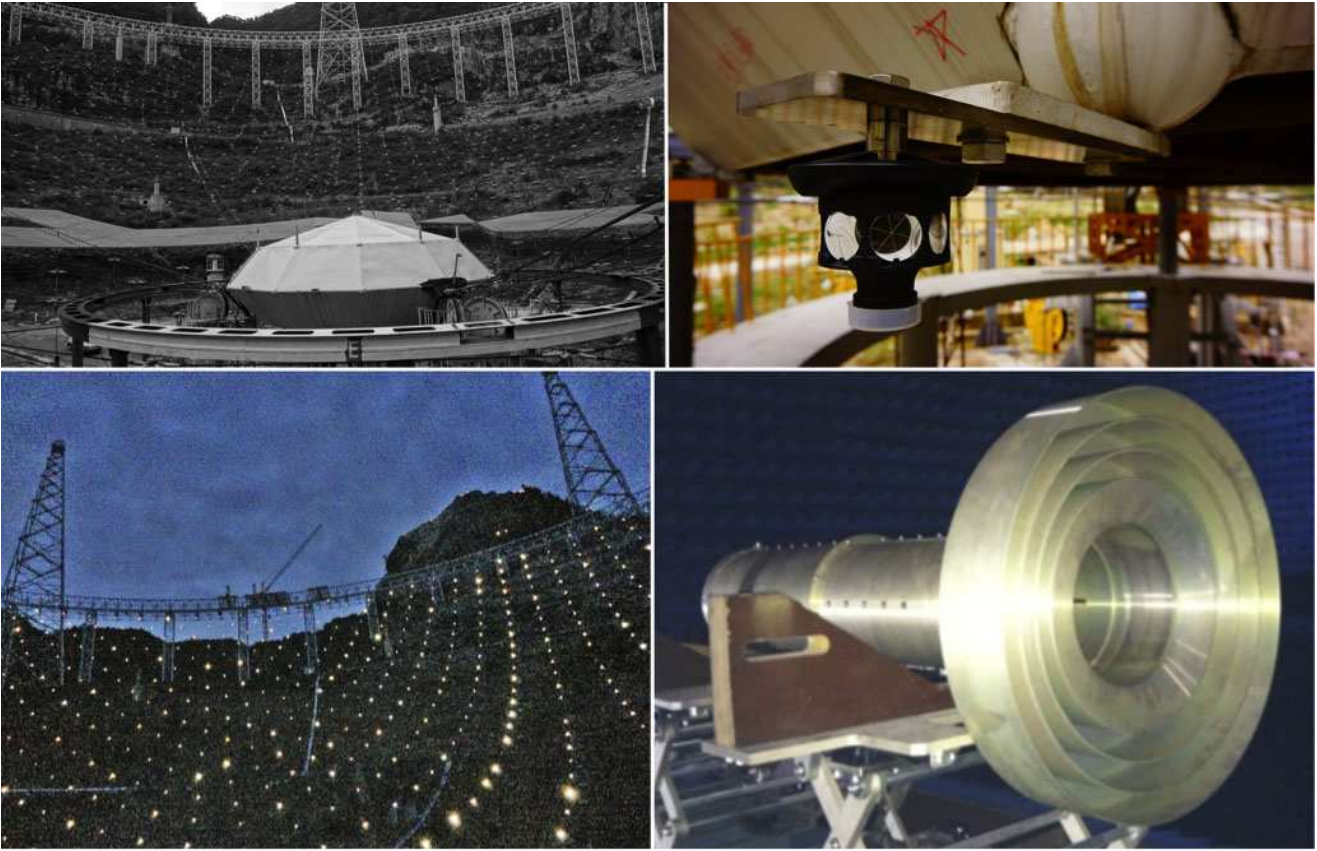}
 \caption{Upper Left: The feed cabin docking station.
 Upper Right: A laser reflector.
 Lower Left: The reflected light from hundreds of reflectors on the cable-mesh system for the primary panels.
 Lower Right: the L-band single pixel feed receiver for FAST.}
   \label{fig2}
\end{center}
\end{figure}

\textbf{3) Feed Cabin Suspension.}
The feed cabin is supported and driven by cables and servomechanisms without any solid structure between the cabin and the towers.
The design team conducted an end-to-end simulation,
which showed that the displacement of the feed cabin following the first adjustment control can be constrained to within a few centimeters, while the secondary stabilizer further reduces the error to a few millimeters.
The test cabin, which is an empty cabin having the same weight as the real cabin, was lifted on the November 21$^{st}$ 2015.
The docking station and the test cabin are shown in Fig. \ref{fig2}.
The actual focal cabin is being assembled on the docking station.
During the normal operation, the cabin can be lowered and parked onto the docking station to allow engineering work around and inside the cabin.

\textbf{4) Measurement and Control.}
The FAST design requires a system-wide positional accuracy of about 1 cm on a range $\gtrsim$130 m,
which amounts to a spatial dynamic range of $\gtrsim$4 orders of magnitude.
Tower stations mounted on 25 measuring towers whose positions are known absolutel, and that are anchored to the bedrocks, will train laser beams onto over 2000 targets to get location and trigonometry measurements in real time.
Most targets are mounted on the connecting nodes of 6 triangular shaped panels.
A complete set of adjustment parameters for reforming the dish can be derived in less than 10 minutes from these measurements and sent to the actuators.
For small distance motion, such as those involved in position switching, the required time will be much less.
A laser reflector and views of the reflected lights from one of the measuring towers are shown in Fig. \ref{fig2}.

\textbf{5) Receiver and Backend.}
FAST will be equipped with seven sets of receivers (see Table 2 in Nan et al. 2011),
covering a frequency range of 70 MHz to 3 GHz.
Here, we will mention only the three main survey instruments.
The 19 beam feed-horn array at L-band will be the primary survey instrument for FAST.
The L-band single pixel receiver (Fig. \ref{fig2}) has been developed in FAST's own laboratory, and covers 1.1 to 1.9 GHz with a return loss better than -20 dB and an isolation better than -40 dB across the band.
A team led by Dr. S. Weinreb of Caltech is developing an ultra-wide band receiver for frequencies between 270 and 1620 MHz.
The ultra-wide band feed will be delivered to the site in June of 2016 and become one of the first instrument to be commissioned.
In a separate paper, we will present the techniques and unique capabilities of using this feed in drift-scan mode for pulsar search.
The radio frequency interference environment is being monitored in view of the fact that we are only able to close down transmitters within 5 km of the telescope site.

There will be two sets of receiver domes.
Switching between receivers within one dome is achieved mechanically on a time scale of minutes.
Switching domes takes at least a full day.

\textbf{6) Observatory.}
The construction of the operations center, located in an adjacent lower depression,
has begun. This will be finished in August 2016 and will provide a usable working area of $\rm \sim$2000 m$^2$.

\section{Key Science Goals}

The key science goals for FAST are based on observables between 70 MHz and 3 GHz,
including the 21 cm HI hyperfine structure line, pulsar emission,
radio continuum, recombination lines, and molecular spectral lines including masers.

The majority of the baryonic matter in the universe is in the form of HI gas.
Compared with the Arecibo Telescope, FAST will have three times the scan speed at L-band and twice the sky coverage.
A key goal for investigation of the Galactic Interstellar Medium (ISM) will be a systematic study of very cold atomic gas,
which better reveals the cold ISM and will be interpreted together with CO surveys of comparable spatial spatial comparable resolution to reveal the conversion of atoms to molecules in our Milky Way.
For the local extragalactic universe, FAST will conduct blind HI surveys to measure gas mass, especially of optically dark galaxies (Giovanelli et al. 2013).
Such a census of gas in the local universe will help explain the discrepancy between dark matter simulations and the observable universe, in particular, the $''$missing baryon problem$''$ (Blumenthal et al. 1986).

Using the 19-beam L-band focal plane array,
FAST is expected to discover over 4000 new pulsars (Smits et al. 2009), about 300 of which should have millisecond spin periods.
Roughly 10\% of these millisecond pulsars are expected to be stable enough to be used in pulsar timing arrays (PTA; e.g. the International Pulsar Timing Array $--$ IPTA, Hobbs et al. 2010).
We expect the contribution of FAST-discovered pulsars
and their subsequent timing to will improve the overall PTA sensitivity by a factor of $\rm \sim$3 for detecting gravitational waves.

FAST will be a powerful instrument for molecular spectroscopy.
Targeted galaxy surveys are expected to increase the sample of currently known extragalactic OH mega-masers
by a factor of 10 with a total of about 1000 detections out to z=2 (Zhang, Li \& Wang 2012).
The sky density of OH mega-masers could provide constraints on the galaxy merger rates in the most active epochs of cosmic star formation.

Although limited to below 3GHz, VLBI (Very Long Baseline Interferometry) will be one of the FAST key science goals.
In a natural weighting, FAST at L-band will improve the EVN sensitivity by about a factor of four and
is still better than the full High Sensitivity Array (HSA) with Arecibo, GBT, phased VLA (Very Large Array), Effelsberg and VLBA (Very Long Baseline Array).
The commissioning of FAST VLBI capabilities will be mainly aided by the Shanghai Astronomical Observatory (SHAO),
which has just finished building the Tianma 65 meter telescope.

\section{Early Science Opportunities}

After completing construction of the telescope, and even before completing astronomical commissioning,
FAST will conduct so-called $''$early science$''$ projects aimed at utilizing its unparalleled sensitivity, and helping accelerate the commissioning and optimization of science operations.

The biggest challenge in FAST's innovative design is achieving the real-time precision control.
The required accuracy increases linearly with the observing frequency.
Long-wavelength science targets will thus be favored for FAST's early-science operations.
The fault rate and limited driving speed of the actuators poses a further challenge,
which may limit the majority of early-science observations to being conducted in drift-scan mode.
An efficient pulsar search and a shallow HI galaxy survey can be conducted commensally.

Our numerical simulations based on an updated version of the pulsar population model of Lorimer et al. (2006)
indicate that the maximum detection rate for a FAST drift-scan pulsar search lies between 400 and 700 MHz (Yue et al. 2013).
Our early focus will be on the local group galaxies, M31 and M33, and the globular clusters associated with our own Galaxy, using the ultra-wide band feed.

We will also carry out a spectral scan of the Orion Nebula,
the most sensitive to date, in the frequency bands below 3 GHz.
A slew of molecules, particularly long-chain carbons (e.g. HC$_x$N, $x=7, 9, 11, \dots$), could be detected.
We are looking into an extension of the Herschel HEXOS (Bergin et al. 2010)
source model for the Orion Nebula to clarify the potential of FAST for the discovery of new spectral lines and new interstellar molecules.

We will also conduct trial shallow surveys of HI in galaxies using the drift-scan mode in preparation for the full survey with the 19-beam feed-horn array.
Combined with ASKAP surveys, a full FAST-ASKAP shallow HI survey promises to deliver over half a million detections (Staveley-Smith 2015),
increasing the total number of known gaseous galaxies by at least an order of magnitude.
Such a study should lay the foundation for future efforts in understanding baryons in the local universe.


%
%
%
%
%
%
%

\begin{acknowledgments}
We thank the support of the State Key Development Program for Basic Research (2012CB821800) of the China Ministry of Science and Technology.
We thank Prof. Zhang, Bo from SHAO for providing the sensitivity estimates of VLBI.
The authors would like to thank the anonymous reviewers for their insightful comments and suggestions that have contributed to improving this paper.
All data for this paper is properly cited and referred to in the reference list.
\end{acknowledgments}

\end{article}
%
%
%
%
%
%

%
%


\end{document}